\documentclass{article}
\usepackage{amssymb}
\usepackage{setspace}
\usepackage{geometry}
\usepackage{graphicx}
\usepackage{slashbox}
\usepackage{times}
\usepackage[ruled,vlined]{algorithm2e}

\author{
Andrei Arion\\
{\small INRIA Futurs--LRI, France}\\
{\small \textsf{Andrei.Arion@inria.fr}}
\and 
Angela Bonifati\\
{\small ICAR CNR, Italy}\\
{\small \textsf{bonifati@icar.cnr.it}}
\and 
Ioana Manolescu\\
{\small INRIA Futurs--LRI, France}\\
{\small \textsf{Ioana.Manolescu@inria.fr}}
\and 
Andrea Pugliese\\
{\small University of Calabria,  Italy}\\
{\small \textsf{apugliese@deis.unical.it}}
}
\newcommand{\join}{{\rhd}\mskip-6mu{\lhd}}
\newcommand{\semijoin}{{\rhd}\mskip-4mu{<}}

\newcommand{\outerjoin}{{\rhd}\mskip-6mu{\lhd}\mskip-4mu{\sqsubset}}

\begin{document}

\title{Path Summaries and Path Partitioning in Modern XML Databases}
\maketitle

\begin{abstract}
We study the applicability of XML path summaries in the context of current-day XML databases. We find that summaries provide an excellent basis for optimizing data access methods, which furthermore mixes very well with path-partitioned stores, and with efficient techniques common in today's XML query processors, such as smart node labels (also known as structural identifiers) and structural joins.

We provide practical algorithms for building and exploiting summaries, and prove its benefits, alone or in conjunction with a path partitioned store, through extensive experiments.
\end{abstract}

~\\
\textbf{Contact author}: Ioana Manolescu\\
INRIA Futurs, Gemo group, 4 rue Jacques Monod, ZAC des Vignes, 91893 Orsay Cedex, France\\
Tel. (+33) 1 72 92 59 20, e-mail: Ioana.Manolescu@inria.fr

\newpage
\section{Introduction}
\label{sec:intro}

Path summaries are classical artifacts for semistructured and XML query processing, dating back to 1997~\cite{goldman97}. From path summaries, the concept of path indexes has derived naturally~\cite{PIndex}: the IDs of all nodes on a given path are clustered together and used as an index. Paths have also been used as a natural unit for organizing not just the index, but the store itself~\cite{XQueCEDBT2004,ToX,XParent,XRel}, and as a support for statistics~\cite{AAN01,paperDataX2004}. 

The state of the art of XML query processing advanced significantly since path summaries were first proposed. Structural element identifiers~\cite{StructJoins,ORDPATH,JayOrdered2002} and structural joins~\cite{StructJoins,HolisticTwigJoins,Grust} are among the most notable new techniques, enabling efficient processing of XML navigation as required by XPath and XQuery.

In this paper, we make the following contributions to the state of the art on path summaries:
\begin{itemize} 
\item We study their size and efficient encoding for a variety of XML document, including very ``hard'' cases for which it has never been considered before. We show summaries are feasible, and useful, even for such extreme cases.
\item We describe an efficient method of static query analysis based on the path summary, enabling a query optimizer to smartly select its data access methods. Similar benefits are provided by a schema, however, summaries apply even in the frequent case when schemas are not available~\cite{LaurentXMLCounting}.
\item We show how to use the result of the static analysis to lift one of the outstanding performance hurdles in the processing of physical plans for structural pattern matching, a crucial operation in XPath and XQuery~\cite{XQuery}: duplicate elimination.
\item We describe time- and space-efficient algorithms, implemented in a freely available library, for building and exploiting summaries. We argue summaries are too useful a technique for a modern XML database system {\em not} to use it.
\end{itemize}
Our XSum library is available for download~\cite{XSum}. It has been successfully used to help manage XML materialized views~\cite{ULoad}, and as a simple GUI in a heterogeneous information retrieval context~\cite{INEX}. We anticipate it will find many other useful applications.

Path summaries have been often investigated in conjunction with path indexes and path-partitioned stores. It is thus legitimate to wonder whether path partitioning is still a valid technique the current XML query processing context~? 

With respect to the path partitioning storage approach, our work makes the following contributions:
\begin{itemize}
\item We show that path partitioning mixes well with recent, efficient structural join algorithms, and in particular enables very selective data access, when used in conjunction with a path summary. 
\item A big performance issue, not tackled by earlier path-partitioned stores~\cite{XParent,LoreIndex,XRel}, concerns document reconstruction, complicated by path fragmentation. We show how  an existing technique for building new XML results can be adapted to this  problem, however, with high memory needs and blocking execution behavior. We propose a new reconstruction technique, and show that it is faster, and most importantly, has an extremely small memory footprint, demonstrating thus the practical effectiveness of path partitioning.
\end{itemize}
This paper is organized as follows. Section~\ref{sec:summaries} presents path summaries and a generic path-partitioned storage model. Section~\ref{sec:associating} tackles efficient static query analysis, based on path summaries. Section~\ref{sec:processing} applies this to efficient query planning, and describes our efficient approach for document reconstruction. Section~\ref{sec:experiments} is our experimental study. We then discuss related works and conclude.

\section{Path summaries and path partitioning}
\label{sec:summaries}
This section introduces XML summaries, and path partitioning.

\begin{figure}
\begin{center}\includegraphics[width=\textwidth]{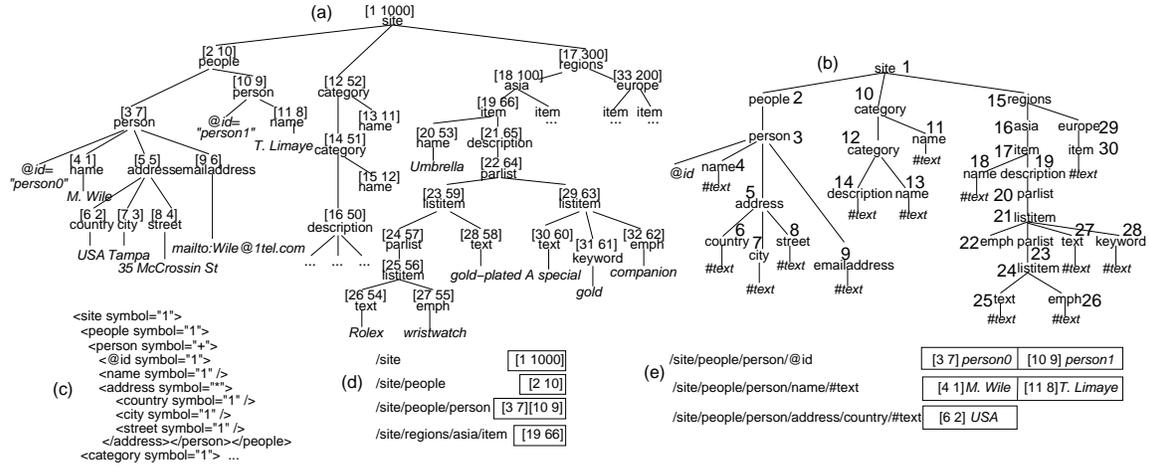}\end{center}
\caption{XMark document snippet, its path summary, and some path-partitioned storage structures.\label{fig:XMark}}
\end{figure}

\subsection{Path summaries}
\label{sec:ps}
The {\em path summary} $PS(D)$ of an XML document $D$ is a tree, whose nodes are labeled with element names from the document. 
The relationship between $D$ and $PS(D)$ can be described based on a function $\phi : D \rightarrow PS(D)$, recursively defined as follows:
\begin{enumerate}
\item $\phi$ maps the root of $D$ into the root of $PS(D)$. The two nodes have the same label.
\item Let $child(n,l)$ be the set of all the $l$-labeled XML elements in $D$, children of the XML element $n$. If $child(n,l)$ is not empty, then $\phi(n)$ has a unique $l$-labeled child $n_l$ in $PS(D)$, and for each 
$n_i \in child(n,l)$, $\phi(n_i)$ is $n_l$.
\item Let $val(n)$ be the set of \#PCDATA children of an element $n \in D$. Then, $\phi(n)$ has an unique child $n_v$ labeled $\#text$, and furthermore, for each $n_i \in val(n)$, $\phi(n_i)=n_v$.
\item Let $att(n,a)$ be the value of the attribute named $a$ of element $n \in D$. Then, $\phi(n)$ has an unique child $n_a$ labeled $@a$, and for each $n_i \in att(n,a)$, we have $\phi(n_i)=n_a$.
\end{enumerate}

Clearly, $\phi$ preserves node labels, and parent-child relationships.
For every simple path $/l_{1}/l_{2}/.../l_{k}$ in $D$, there is exactly one
node reachable by the same path in $PS(D)$. Conversely, each node in $PS(D)$ corresponds to a simple path in $D$.

Figure~\ref{fig:XMark}(b) shows the path summary for the XML fragment at its left. Path numbers appear in large font next to the summary nodes.

We add to the path summary some more information, conceptually related to schema constraints. More precisely, for any summary nodes $x$, $y$ such that $y$ is a child of $x$, we record on the edge $x$-$y$ whether every node on path $x$ has {\em exactly one child} on path $y$, or {\em at least one child} on path $y$, or may lack $y$ children. This information is used for query optimization, as Section~\ref{sec:associating} will show.

\paragraph*{Direct encoding} Let $x$ be the parent of $y$ in the path summary. A simple way to encode the above information is to annotate $y$, the child node, with: \textsf{\small 1} {\em iff} every node on path $x$ has exactly one child on path $y$; \textsf{\small +} {\em iff} every node on path $x$ has at least one child on path $y$, and some node on path $x$ has several children on path $y$. This encoding is simple and compact. However, if we need to know how many {\em descendents} on path $z$ can a node on path $x$ have, we need to inspect the annotations of all summary nodes between $x$ and $z$.

\paragraph*{Pre-computed encoding} Starting from the \textsf{\small 1} and \textsf{\small +} labels, we compute more refined information, which is then stored in the summary, while the \textsf{\small 1} and \textsf{\small +}  labels are discarded. 

We identify clusters of summary nodes connected between them only with \textsf{\small 1}-labeled edges; such clusters form a {\em 1-partition} of the summary. Every cluster of the 1-partition is assigned a \textsf{\small n1} label, and this label is added to the serialization of every path summary node belonging to that cluster. Then, a node on path $x$ has exactly one descendent on path $z$ {\em iff} $x$.\textsf{\small n1}$=$$z$.\textsf{\small n1}. We also build a {\em +-partition} of the summary, aggregating the 1-partition clusters connected among them only by \textsf{\small +} edges, and similarly produce \textsf{\small n+} labels, which allow to decide whether nodes on path $x$ have at least one descendent on path $z$ by checking whether $x$.\textsf{\small n+}$=$$z$.\textsf{\small n+}.

\paragraph*{Building and storing summaries} For a given document, let $N$  denote its size, $h$ its height, and $|PS|$ the number of nodes in its path summary. In the worst case, $|PS|=N$, however our analysis in Table~\ref{tab:PS} demonstrates that this is not the case in practice. The documents in Table~\ref{tab:PS} are obtained from~\cite{UWXML}, except for the XMark$n$ documents, which are generated~\cite{XMark} to the size of $n$ MB, and two DBLP snapshots from 2002 and 2005~\cite{DBLP}.

A first remark is that for all but the TreeBank document, the summary has at most a few hundreds of nodes, and is 3 to 5 orders of magnitude smaller than the document. A second remark is that as the XMark and DBLP documents grow in size, their respective summaries grow very little. Intuitively, the structural complexity of a document tends to level up, even if more data is added, even for complex documents such as XMark, with 12 levels of nesting, recursion etc.
A third remark is that TreeBank, although not the biggest document, has the largest summary (also, the largest we could find for real-life data sets). TreeBank is obtained from natural language, into which tags were inserted to isolate parts of speech. While we believe such documents are rare, robust algorithms for handling such summaries are needed, if path summaries are to be included in XML databases.

\begin{table*}[t!]
\begin{center}
\begin{tabular}{||c|r|r|r|r||}
\hline
\hline
Doc. & Shakespeare &  Nasa & Treebank & SwissProt \\
\hline
\hline
Size & 7.5 MB & 24 MB & 82 MB & 109 MB \\
\hline
$N$ & 179,690 & 476,645 & 2,437,665 & 2,977,030\\
\hline
$|PS|$ & 58 & 24 & 338,738 & 117 \\
\hline
$|PS|/N$ & 3.2*10$^{-4}$  & 5.0*10$^{-5}$ & 1.3*10$^{-1}$  & 3.9*10$^{-5}$ \\
\hline
\hline
\end{tabular}

~\\

\begin{tabular}{||c|r|r|r|r|r||}
\hline
\hline
Doc. & XMark$11$ & XMark$111$ & XMark$233$ & DBLP (2002) & DBLP (2005) \\
\hline
\hline
Size & 11 MB & 111 MB & 233 Mb & 133 MB & 280 MB \\
\hline
$N$ & 206,130 & 1,666,310 & 4,103,208 & 3,736,406 & 7,123,198 \\
\hline
$|PS|$ & 536 & 548 & 548 & 145 & 159\\
\hline
$|PS|/N$ & 2.4*10$^{-3}$ & 3*10$^{-4}$ & 1.3*10$^{-4}$  & 3.8*10$^{-5}$& 2.2*10$^{-5}$ \\
\hline
\end{tabular}
\caption{Sample XML documents and their path summaries.\label{tab:PS}}
\end{center}
\end{table*}

A path summary is built during a single traversal of the document, in $O(N)$ time, using $O(|PS|)$ memory~\cite{AAN01,goldman97}. Our implementation  gathers \textsf{\small 1} and \textsf{\small +} labels during summary construction, and traverses the summary again if the pre-computed encoding is used, making for $O(N+|PS|)$ time and $O(|PS|)$ memory. 
This linear scaleup is confirmed by the following measures, where the summary building times $t$ are scaled to the time for XMark$11$:

\begin{center}
\begin{tabular}{|c|r|r|r|r|}
\hline
XMark$n$&XMark$2$&XMark$11$&XMark$111$&XMark$233$\\
\hline
$n/11$& 0.20 & 1.0 & 9.98 & 20.02 \\
\hline
$t/t_{11\mbox{ Mb}}$& 0.32 & 1.0 & 8.58 & 15.84 \\
\hline 
\end{tabular}
\end{center}

Once constructed, a summary must be stored for subsequent use. To preserve the summary's internal structure, we will store it as a tree,  leading to $O(|PS|)$ space occupancy if the basic encoding is used, and $O(|PS|*log_2|PS|)$ if the pre-computed encoding is used (\textsf{\small n1} and \textsf{\small n+} labels grow in the worst case up to $N$). We evaluate several summary serialization strategies in Section~\ref{sec:experiments}.

\subsection{Path-partitioned storage model}
\label{sec:pp}
{\em Structural identifiers} are assigned to each element in an XML document. A direct comparison of two structural identifiers suffices to decide whether the corresponding elements are structurally related (one is a parent or ancestor of the other) or not. A very popular such scheme consists of assigning \textsf{\small (pre,post,depth)} numbers to every node~\cite{StructJoins,EfficientStructJoins,MixedMode}.
The \textsf{\small pre} number corresponds to the positional number of the element's begin tag, and the \textsf{\small post} number corresponds to the number of its end tag in the document.  For example,  Figure~\ref{fig:XMark}(a) depicts \textsf{\small (pre,post)} IDs  above the elements. The \textsf{\small depth} number reflects the element's depth in the document tree (omitted in Figure~\ref{fig:XMark} to avoid clutter).
Many variations on the \textsf{\small (pre,post,depth)} scheme exist, and more advanced structural IDs have been proposed, such as DeweyIDs~\cite{JayOrdered2002} or ORDPATHs~\cite{ORDPATH}.
While we use \textsf{\small (pre,post)} for illustration, the reader is invited to keep in mind that any structural ID scheme can be used.

~\\
Based on structural IDs, our first structure contains a compact representation of the XML tree structure. We \emph{partition the identifiers according to the data path} of the elements. For each path, we create an {\em ID path sequence}, which is the sequence of IDs in document order.
Figure~\ref{fig:XMark}(d) depicts a few ID path sequences resulting from some paths of the sample document in Figure~\ref{fig:XMark}(a).

~\\
Our second structure stores the contents of XML elements, and values of the attributes.
We pair such values to an ID of their closest enclosing element identifier. 
Figure~\ref{fig:XMark}(e) shows some such (ID, value) pair sequences for our sample document.

\section{Computing paths relevant to query nodes}
\label{sec:associating}
An important task of a query optimizer is {\em access method selection}: given a set of stored data structures (such as base relations, indexes, or materialized views) and a query, find the data structures which may include the data that the query needs to access. An efficient access method selection process requires:

\begin{itemize}
\item a store providing selective access methods;
\item an optimizer able to correctly identify such methods.
\end{itemize}

The main observation underlying this work is that path summaries provide very good support for the latter; we explain the principle in Section~\ref{sec:idea} and provide efficient algorithms supporting it in Section~\ref{sec:computingPaths}.  A path-partitioned storage, moreover, provides robust and selective data access methods (see Section~\ref{sec:processing}).

\begin{figure}
\begin{center}\includegraphics[width=\textwidth]{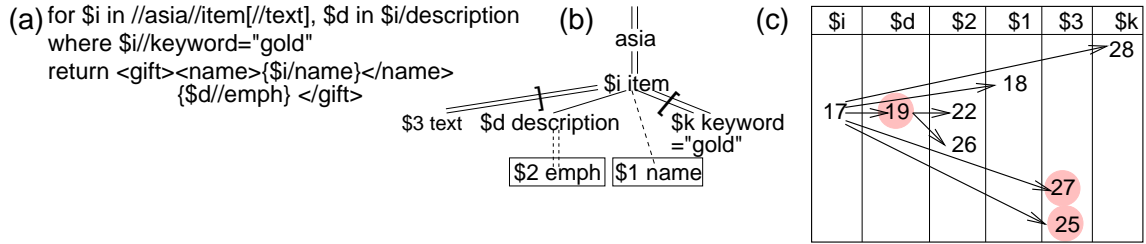}
\caption{(a): sample query; (b):~resulting query pattern; (c):~resulting paths on the document in Figure~\ref{fig:XMark}.\label{fig:GTP}}
\end{center}
\end{figure}

\subsection{The main idea}
\label{sec:idea}
Given an XQuery query $q$, the optimizer must identify all data structures containing information about any XML node $n$ that must be accessed by the execution engine when evaluating $q$. In practice, the goal is to identify structures containing {\em a tight superset} of the data strictly needed, given that the storage usually does not contain a materialized view for any possible query.

Paths provide a way of specifying quite tight supersets of the nodes that query evaluation needs to visit. 

For instance, for the query \textsf{\small //asia//item[description]/name}, given the summary in Figure~\ref{fig:XMark}, elements on paths 17 must be returned, therefore data from paths 18 to 28 may need to be retrieved. Query evaluation does not need to inspect elements from other paths. For instance, paths 4 and 11 are not relevant for the query, even though they correspond to \textsf{\small name} elements; similarly, \textsf{\small item} elements on path 30 are not relevant. 
These examples illustrate how ancestor paths, such as \textsf{\small //asia} (16) filter descendent paths, separating 17 (relevant) from 30 (irrelevant). Descendent paths can also filter ancestor paths. For instance, DBLP contains \textsf{\small article}, \textsf{\small journal}, \textsf{\small book} elements etc. The query \textsf{\small //*[inproceedings]} must access \textsf{\small /dblp/article} elements, but it does not need to access \textsf{\small /dblp/journal} or \textsf{\small /dblp/book} elements, since they never have \textsf{\small inproceedings} children.

~\\
Let us consider the process of gathering, based on a path summary, the relevant data paths  for a query. We consider the downward, conjunctive XQuery subset from~\cite{GTP}. Every query yields a query pattern in the style of~\cite{GTP}. Figure~\ref{fig:GTP} depicts an XQuery query (a), and its pattern (b). We distinguish parent-child edges (single lines) from ancestor-descendent ones (double lines). Dashed edges represent optional relationships: the children (resp. descendents) at the lower end of the edge are not required for an element to match the upper end of the edge. Edges crossed by a ``[``  connect parent nodes with children that  must be found in the data, but are not returned by the query, corresponding to navigation steps in path predicates, and in ``where'' XQuery clauses.
We call such nodes {\em existential}.
Boxed nodes are those which must actually be returned by the query. 
In Figure~\ref{fig:GTP}(b), some auxiliary variables \textsf{\small \$1}, \textsf{\small \$2} and \textsf{\small \$3} are introduced for the expressions in the return clause, and expressions enclosed in  existential brackets $[\;]$.

For every node in the pattern, we compute a {\em minimal set of relevant paths}.  A path $p$ is relevant for node $n$ {\em iff}: ($i$)~the last tag in $p$ agrees with the tag of $n$ (which may also be *); ($ii$)~$p$ satisfies the structural conditions imposed by the $n$'s ancestors, and ($iii$)~$p$ has descendents paths in the path summary, matching all non-optional descendents of the node. 
Relevant path sets are organized in a tree structure, mirroring the relationships between the nodes to which they are relevant in the pattern. 

The paths relevant to nodes of the pattern in Figure~\ref{fig:GTP}(b) appear in Figure~\ref{fig:GTP}(c).  The paths surrounded by grey dots are relevant, but not part of the minimal relevant sets, since they are either {\em useless ``for'' variable paths}, or {\em trivial existential node paths}.

\paragraph*{Useless ``for'' variable path} The path 19 for the variable \textsf{\small \$d}, although it satisfies condition 1, has no impact on the query result, on a document described by the path summary in Figure~\ref{fig:XMark}. This is because: ($i$)~\textsf{\small \$d} is not required to compute the query result; ($ii$)~it follows from the path summary that every element on path 17 (relevant for \textsf{\small \$i}) has exactly one child on path 19 (relevant to \textsf{\small \$d}). This can be seen by checking that 19 is annotated with a \textsf{\small 1} symbol. Thus, query evaluation does not need to find bindings for \textsf{\small \$d}. Instead, it suffices to bind \textsf{\small \$i} and \textsf{\small \$2} to the correct paths and combine them, shortcircuiting the binding of \textsf{\small \$d}.

In general, a path $p_x$ relevant for a ``for'' variable \textsf{\small \$x} is useless as soon as the following two conditions are met:
\begin{enumerate}  
\item \textsf{\small \$x}, or path expressions starting from \textsf{\small \$x}, do not appear in a ``return'' clause.
\item If \textsf{\small \$x} has a parent \textsf{\small \$y} in the query pattern, let $p_y$ be the path relevant for \textsf{\small \$y}, ancestor of $p_x$. Then, all summary nodes on the path from some child of $p_y$, down to $p_x$, must be annotated with the symbol \textsf{\small 1}. If, on the contrary, \textsf{\small \$x} does not have a parent in the query pattern, then all nodes from the root of the path summary to $p_x$ must be annotated with \textsf{\small 1}. 
\end{enumerate}

Such a useless path $p_x$ is erased from its path set.
If \textsf{\small \$x} had a parent \textsf{\small \$y} in the pattern, then there exists a path $p_y$, ancestor of $p_x$, relevant for \textsf{\small \$y}. If \textsf{\small \$x} has some child \textsf{\small \$z} in the pattern, in the final solution, an arrow will point directly from $p_y$ to the paths relevant for $p_z$, shortcircuiting $p_x$. 
In Figure~\ref{fig:GTP}, once 19 is found useless, 17 will point directly to the paths 22 and 26 in the relevant set for \textsf{\small \$2}.

\paragraph*{Trivial existential node paths} 
The path summary in Figure~\ref{fig:XMark} guarantees that every XML element on path 17 has at least one descendent on path 27. This is shown by the \textsf{\small 1} or \textsf{\small +} annotations on all paths between 17 and 27.
In this case, we say 27 is a trivial path for the existential node \textsf{\small \$3}. If the annotations between 17 and 25 are also \textsf{\small 1} or \textsf{\small +}, path 25 is also trivial. The execution engine does not need to check, on the actual data, which elements on path 17 actually have descendents on paths 25 and 27: we know they all do. Thus, paths 25 and 27 are discarded from the set of \textbf{\small \$3}. 

In general, let $p_x$ be a path relevant for an existential node \textsf{\small \$x}; this node must have had a parent or ancestor \textsf{\small \$y} in the pattern, such that the edge going down from \textsf{\small \$y}, on the path connecting \textsf{\small \$y} to \textsf{\small \$x}, was marked by a ``[``. There must be a path $p_y$ relevant for \textsf{\small \$y}, such that $p_y$ is an ancestor of $p_x$. We say $p_x$ is a trivial path if the following conditions hold:

\begin{enumerate}
\item All summary nodes between $y$ and $x$ are annotated with either \textsf{\small 1} or \textsf\small{+}.
\item All paths descendent of $p_x$, and relevant for nodes below \textsf{\small \$x} in the query pattern, are trivial. 
\item No value predicate is applied on \textsf{\small \$x} or its descendents.
\end{enumerate}

After pruning out useless and trivial paths, nodes left without any relevant path are eliminated; the connecteed paths of the remaining nodes are returned.
For the query pattern in Figure~\ref{fig:GTP}(b), this yields exactly the result in Figure~\ref{fig:GTP}(c) from which the grey-dotted paths, and their pattern nodes, have been erased.

\subsection{Computing relevant paths}
\label{sec:computingPaths}
Having defined minimal sets of relevant paths, the question is how to efficiently compute them. This problem has not been tackled before.
Moreover, trivial algorithms (such as string matching of paths) do not apply, due to the complex tree structure of query patterns, and to the tree-structured connections between relevant paths. Such methods cannot minimize path sets, either.

A straightforward method is a recursive parallel traversal of $PS$ and $q$, checking ancestor conditions for a path to be relevant for a pattern node during the descent in the traversal. When a path $p$ satisfies the ancestor conditions for a pattern node $n$, the summary subtree rooted in $p_n$ is checked for descendent paths corresponding to the required children of $n$ in the pattern. This has the drawback of visiting a summary node more than once. For instance, consider the query \textsf{\small //asia//parlist//listitem}: on the summary in Figure~\ref{fig:XMark}, the subtree rooted at path 24 will be traversed once to check descendents of path 20, and once to check descendents of the path 23. 

A more efficient method consists of performing a single traversal of the summary, and collecting potentially relevant paths, which satisfy the ancestor path constraints, but not necessarily (yet) the descendent path constraints. When the summary subtree rooted at a potentially relevant path has been fully explored, we  check if the required descendent paths have been found during the exploration. Summary node annotations are also collected during the same traversal, to enable identification of useless and trivial paths.

\begin{figure}[t!]
\begin{center}
\includegraphics[width=7.3cm]{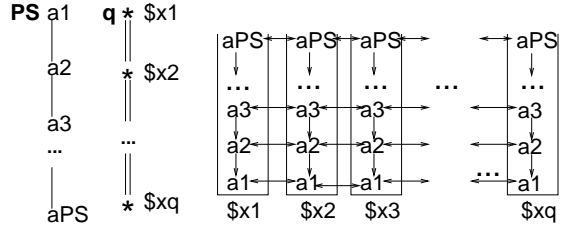}
\caption{Sample query pattern and relevant path sets.\label{fig:bounds}}
\end{center}
\end{figure}

The total size of the relevant path sets may be quite important, as illustrated in Figure~\ref{fig:bounds}. Here, any subset of $|q|$ nodes of $PS$ contains one path relevant for every node of $q$, leading to a cumulated size of  $|q|!*(|PS|-|q|)!/|PS|!$ relevant paths. This is problematic with large summaries: relevant path identification is just an optimization step, and should not consume too much memory, especially in a multi-user, multi-document database. Therefore, a compact encoding of relevant path sets is needed.

The single-traversal algorithm described above may run on an in-memory de-serialized summary. A more efficient alternative is to traverse the summary in streaming fashion, using only $O(h)$ memory to store the state of the traversal. The algorithm we propose to that effect is shown in Algorithm~\ref{algo:pathmatch}; it runs in two phases.

\incmargin{1em}
\linesnumbered
\begin{algorithm}[t]
\SetKwInOut{Input}{Input}
\SetKwInOut{Output}{Output}
\dontprintsemicolon
\label{algo:pathmatch}
\caption{Finding minimal relevant path sets}
\Input{query pattern $q$}
\Output{the minimal set of relevant paths $paths(n)$ for each pattern node $n$}

\tcc{\textbf{\hfill\hfill Phase 1: finding relevant paths \hfill}}
\tcc{ Create one stack for each pattern node:}
\ForEach{pattern node $n$}{{\em stacks(n)} $\leftarrow$ new stack}\;
currentPath $\leftarrow 0$\;
Traverse the path summary in depth-first order:\\
\ForEach{node $n$ visited for the first time}{Run algorithm \textbf{beginSummaryNode}}
\ForEach{node $n$ whose exploration is finished}{Run algorithm \textbf{endSummaryNode}}

\tcc{\textbf{\hfill\hfill Phase 2: minimizing relevant path sets\hfill}}

\ForEach{node $n$ in $q$}{
  \ForEach{stack entry $se$ in stacks(n)}{
      {\If{$n$ is existential and \\\textbf{all1or+}(se.parent.path, se.path)}{
      $se$ is trivial. Erase $se$ and its descendants from the stack.}}
  {\If{$n$ is a ``for'' var. and $n$ and its desc. are not boxed and \textbf{all1}(se.parent.path,se.path)}{
	$se$ is useless. Erase $se$ from {\em stacks(n) and }\; connect $se$'s parent to $se$'s children, if any}}}
  {\em paths(n)}$\leftarrow$ paths in all remaining entries in {\em stacks(n)}
}
\end{algorithm}
\decmargin{1em}

\incmargin{1em}
\linesnumbered
{\small \begin{algorithm}[t]
\SetKwInOut{Input}{Input}
\dontprintsemicolon
\label{algo:beginPSNode}
\caption{\textbf{beginSummaryNode}}
\Input{current path summary node labeled $t$}
\tcc{ Uses the shared variables {\em currentPath}, {\em stacks} }
currentPath $++$;\;
\tcc{ Look for pattern query nodes which $t$ may match:}
\ForEach{pattern node $n$ s.t. $t$ matches $n$'s label}{
\tcc{ Check if the current path is found in the correct context {\em wrt} $n$:}
  {\If{(1)~$n$ is the topmost node in $q$, or (2)~$n$ has a parent node $n'$, {\em stacks(n')} is not empty, and {\em stacks(n').top} is open}{
  {\If{the level of currentPath agrees with the edge above $n$, and with the level of {\em stacks(n').top}}{
  \tcc{ The current path may be relevant for $n$, so create a candidate entry for {\em stacks(n):}}
  stack entry {\em se} $\leftarrow$ new entry(currentPath)\;
  {\em se.parent} $\leftarrow$ {\em stacks(n').top}\;
  {\If{stacks(n) is not empty and stacks(n).top is open}{
      {\em se.selfParent} $\leftarrow$ {\em stacks(n).top}}\Else{{\em se.selfParent} $\leftarrow$ {\em null}}}
  {\em se.open} $\leftarrow$ true\;
  {\em stacks(n).push(se)}\;
}}
}}
} 
\end{algorithm}}

\incmargin{1em}
\linesnumbered
\begin{algorithm}[h]
\SetKwInOut{Input}{Input}
\dontprintsemicolon
\label{algo:endPSNode}
\caption{\textbf{endSummaryNode}}
\Input{current path (node in the path summary), labeled $t$}
\tcc{ Uses the shared variables currentPath, {\em stack}}
\ForEach{query pattern node $n$ s.t. {\em stacks(n)} contains an entry $se$ for currentPath}{
  \tcc{ Check if {\em currentPath} has descendents in the stacks of non-optional $n$ children:}
  \ForEach{non-optional child $n'$ of $n$}{
    {\If{$se$ has no children in stacks(n')}{
	{\If{$se.ownParent \neq null$}{
	    connect $se$ children to $se.ownParent$\;
	    pop $se$ from {\em stacks(n)}\;}
	 \Else{
	    pop $se$ from {\em stacks(n)}\;
	    pop all $se$ descendent entries from their stack\;
	}}
}}
    }
 $se.open \leftarrow false$\;
}
\end{algorithm}

\textbf{Phase 1 (finding relevant paths)} performs a streaming traversal of the summary, and applies Algorithm~\ref{algo:beginPSNode} whenever entering a summary node, and Algorithm~\ref{algo:endPSNode} when leaving the node. Algorithm~\ref{algo:pathmatch} uses one stack for every pattern node, denoted $stack(n)$. Potentially relevant paths are gathered in stacks, and eliminated when they are found irrelevant, useless or trivial. 
An entry in $stacks(n)$ consists of:

\begin{itemize}
\item A {\em path} (in fact, the path number).
\item A {\em parent} pointer to an entry in the stack of $n$'s parent, if $n$ has a parent in the pattern, and {\em null} otherwise.
\item A {\em selfparent} pointer. This points to a previous entry on the same stack, if that entry's path number is an ancestor of this one's, or {\em null} if such an ancestor does not exist at the time when the entry has been pushed.
Self-pointers allow to compactly encode relevant path sets.
\item An {\em open} flag. This is set to {\em true} when the entry is pushed, and to {\em false} when all descendents of $p$ have been read from the path summary. Notice that we cannot afford to pop the entry altogether when it is no longer open, since we may need it for further checks in Algorithm~\ref{algo:endPSNode} (see below).
\item A set of {\em children} pointers to entries in $n$'s children's stacks.
\end{itemize}

Figure~\ref{fig:bounds} outlines the content of all stacks after relevant path sets have been computed for $q$. Horizontal arrows between stack entries represent {\em parent} end {\em children} pointers; downward vertical arrows represent {\em selfparent} pointers, which we explain shortly. 



In Algorithm \textbf{beginSummaryNode}, when a summary node (say $p$) labeled $t$ starts, we need to identify pattern query nodes $n$ for which $p$ may be relevant.
A first necessary condition concerns the final tag in $p$: it must be $t$ or $*$ in order to match a $t$-labeled query node. 
A second necessary condition concerns the {\em context} in which $p$ is encountered: at the time when traversal enters $p$, there must be an open, potentially relevant path for $n$'s parent is an ancestor of $p$.  This can be checked by verifying that there is an entry on the stack of $n'$, and that this entry is $open$.
If $n$ is the top node in the pattern, if it should be a direct child of the root, then so should $p$.
If both conditions are met, an entry is created for $p$, and connected to its parent entry (lines 5-6). 

The {\em selfparent} pointers, set at the lines 7-10 of Algorithm~\ref{algo:beginPSNode}, allow sharing children pointers among entries nodes in the same stack. 
For instance, in the relevant node sets in Figure~\ref{fig:bounds}, node \textsf{\small a1} in the stack of \textsf{\small \$x1} only points to \textsf{\small a1} in the stack of \textsf{\small \$x2}, even though it should point also to nodes \textsf{\small a2}, \textsf{\small a3}, $\ldots$, \textsf{\small aPS} in the stack of \textsf{\small \$x2}, given that these paths are also in descendent-or-self relationships with \textsf{\small a1}. 
The information that these paths are children of the \textsf{\small a1} entry in the stack of \textsf{\small \$x1} is implicitly encoded by the {\em selfparent} pointers of nodes further up in the \textsf{\small \$x1} stack: if path \textsf{\small a3} is a descendent of the \textsf{\small a2} entry in this stack, the \textsf{\small a3} is implicitly a descendent of the \textsf{\small a1} entry also.

This stack encoding via {\em selfparent} is inspired from the Holistic Twig Join~\cite{HolisticTwigJoins}. The differences are: ($i$)~we use it when performing a single streaming traversal over the summary, as opposed to joining separate disk-resident ID collections; ($ii$)~we use it on the summary, at a smaller scale, not on the data itself. However, as we show in Section~\ref{sec:experiments}, this encoding significantly reduces space consumption in the presence of large summaries. This is important, since  real-life systems are not willing to spend significant resources for optimization.
In  Figure~\ref{fig:bounds}, based on {\em selfparent}, the relevant paths are encoded in only $O(|q|*|PS|)$. Our experimental evaluation in Section~\ref{sec:experiments} shows that this upper bound is very relaxed.

In line 11 of Algorithm~\ref{algo:beginPSNode}, the new entry $se$ is marked as {\em open}, to signal that subsequent matches for children of $n$ are welcome, and pushed in the stack. 

Algorithm \textbf{endSummaryNode}, before finishing the exploration of a summary node $p$, checks and may decide to erase the stack entries generated from $p$.  A stack entry is built with $p$ for a node $n$, when $p$ has all the required {\em ancestors}. However, \textbf{endSummaryNode} still has to check whether $p$ had all the required {\em descendents}.
Entry $se$ must have at least one {\em child} pointer towards the stacks of all required children of $n$; otherwise, $se$ is not relevant and is discarded. In this case, its  descendent entries in other stacks are also discarded, if these entries are not indirectly connected (via a {\em selfparent} pointer) to an ancestor of $se$. If they are, then we connect them directly to {\em se.selfparent}, and discard only $se$  (lines 4-9).

The successive calls to \textbf{beginPathSummaryNode} and \textbf{endPathSummaryNode} lead to entries being pushed on the stacks of each query node. Some of these entries left on the stacks may be trivial or useless; we were not able to discard them earlier, because they served as ``witnesses'' that validate their parent entries (check performed by Algorithm~\ref{algo:endPSNode}).

\textbf{Phase 2 (minimizing relevant path sets)} in Algorithm~\ref{algo:pathmatch} goes over the relevant sets and prunes out the trivial and useless entries. 
The predicate \textbf{all1($p_x,p_y$)} returns true if all nodes between $p_x$ and $p_y$ in the path summary are annotated with \textsf{\small 1}. Similarly, \textbf{all1or+} checks if the symbols are either \textsf{\small 1} or \textsf{\small +}. Useless entries are ``short-circuited'', just like Algorithm~\ref{algo:endPSNode} did for irrelevant entries. At the end of this phase, the entries left on the stack are the minimal relevant path set for the respective node.

Evaluating \textbf{all1} and \textbf{all1or+} takes constant time if the pre-computed encoding is used (Section~\ref{sec:summaries}). With the basic encoding, Phase 2 is actually a second summary traversal (although for readability, Algorithm~\ref{algo:pathmatch} does not show it this way). For every $p_x$ and $p_y$ such that Phase 2 requires evaluating \textbf{all1($p_x,p_y$)} and \textbf{all1or+($p_x,p_y$)}, the second summary traversal verifies the annotations of paths from $p_x$ to $p_y$, using constant memory. 

\paragraph*{Overall time and space complexity} The time complexity of Algorithm~\ref{algo:pathmatch} depends linearly on $|PS|$.
For each path, some operations are performed for each query pattern node for which the path may be relevant. In the worst case, this means a factor of $|q|$. 
The most expensive among these operations, is checking that an entry had at least one child in a set of stacks. If we cluster an entry's children by their stack, this takes at most $|q|$ steps.
Putting these together, we obtain $O(|PS|*|q|^2)$ time complexity.
The space complexity in $O(|PS|*|q|)$ for encoding the path sets.

\section{Query planning and processing based on relevant path sets}
\label{sec:processing}
We have shown how to obtain for every query pattern node $n$, a set of relevant paths {\em paths(n)}. 

No matter which particular fragmentation model is used in the store, it is also possible to compute the paths associated to every storage structure, view, or index. For example, assume a simple collection of structural identifiers for all elements in the document, such as the Element table in~\cite{XRel} or the basic table considered in~\cite{Grust}: the path set associated to such a structure includes all $PS$ paths. As another example, consider an index grouping structural IDs by the element tags, as in~\cite{TimberVLDBJ} or the LIndex in~\cite{LoreIndex}: the path set associated to every index entry includes all paths ending in a given tag. A path index such as PIndex~\cite{LoreIndex} or a path-partitioned store~\cite{ToX} provides access to data from one path at a time. 

Based on this observation, we recommend the following simple access path selection strategy:

\begin{itemize}
\item Compute relevant paths for query pattern nodes.
\item Compute associated paths for data in every storage structure (table, view, index etc.)
\item Choose, for every query pattern node, a storage structure whose associated paths form a (tight) superset of the node's relevant paths.
\end{itemize}

This general strategy can be fitted to many different storage models. Section~\ref{sec:planning} explores it for the particular case of a path-partitioned storage model. Section~\ref{sec:std} and~\ref{sec:reconstruct} show how path information may simplify the physical algorithms needed for structural join processing, respectively, complex output construction.

\subsection{Constructing query plans on a path-partitioned store}
\label{sec:planning}
\begin{figure}
\begin{center}
\includegraphics[width=0.5\columnwidth]{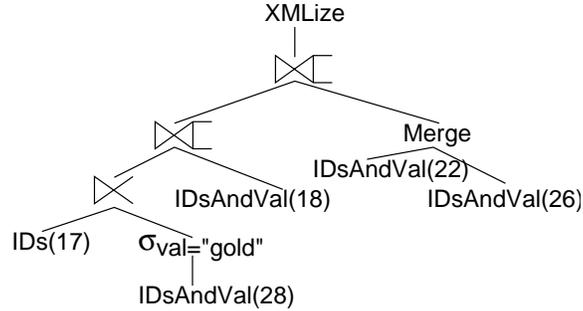}
\caption{Complete QEP for the query in Figure~\ref{fig:GTP}.\label{fig:finalQEP}}
\end{center}
\end{figure}

With a path-partitioned store, IDs and/or values from every path are individually accessible. In this case, the general access method selection approach becomes: ($i$)~construct access plans for every query pattern node, by merging the corresponding ID or value sequences (recall the logical storage model from Figure~\ref{fig:XMark}); ($ii$)~combine such access plans as required by the query, via structural joins, semijoins, and outerjoins. To build a complete query plan (QEP), the remaining steps are: ($iii$)~for every relevant path $p_{ret}$ of an expression appearing in a ``return'' clause, reconstruct the subtrees rooted on path $p_{ret}$; ($iv$)~re-assemble the output subtrees in the new elements returned by the query. For example, Figure~\ref{fig:finalQEP} depicts a QEP for the sample query from Figure~\ref{fig:GTP}. In this QEP, IDs($n$) designates an access to the sequence of structural IDs on path $n$, while IDAndVal($n$) accesses the (ID, value) pairs where IDs identify elements on path $n$, and values are text children of such elements. The left semi-join ($\semijoin$) and the left outer-joins ($\outerjoin$) are {\em structural}, i.e. they combine inputs based on parent-child or ancestor-descendent relationships between the IDs they contain. Many efficient algorithms for structural join exist~\cite{StructJoins,HolisticTwigJoins}; we are only concerned here with the logical operator.

The plan in Figure~\ref{fig:finalQEP} is directly derived from the relevant path sets, shown at the end of Section~\ref{sec:idea}, and the query itself. The selection $\sigma$ has been taken from the query, while the Merge fuses the information from the two relevant paths for \textsf{\small \$2} (\textsf{\small emph} elements). The final XMLize operator assembles the pieces of data in a result. Section~\ref{sec:reconstruct} studies this in more details.

The QEP in Figure~\ref{fig:finalQEP} takes good advantage of relevant paths to access only a very small subset of the data present in an XMark document. For instance, only asian item data is read (not from all items), and only the useful fragments of this data; for instance, voluminous item \textsf{\small descriptions} do not need to be read. 

Clearly, more elaborate index structures such as the F\&B index~\cite{FBIndex} may lead to accessing even less data, providing e.g. direct access only to those \textsf{\small item}s (path 17) that have \textsf{\small keyword}s (path 28). However, even an F\&B index cannot provide \textsf{\small item} IDs with their {\em optional} \textsf{\small emph} children (paths 22 and 26), because we are interested also in items that {\em do not} have such children, while the F\&B index and its variants correspond to {\em conjunctive} paths only (all items are required). Furthermore, the selection $\sigma$ still needs to be applied on on \textsf{\small keyword}s, and path indexes cannot help here. 

The conclusion we draw is: while ellaborate structure indexing schemes can cover more complex path patterns, the simplicity and robustness of a simple path-driven access to IDs, combined with the ability to do {\em most of query processing just by efficient structural ID joins}, provide good support for efficient query evaluation in current-day XML database engines.

\subsection{Using path annotations for efficient structural joins plans}
\label{sec:std}
Path expressions used in XPath and XQuery need to return duplicate-free lists of nodes, in specific orders. Structural joins, in contrast, may introduce spurious duplicates, whose elimination is expensive.
Schema-based techniques have been developed to decide when duplicate elimination is unnecessary~\cite{michiels}. However, schemas are often unavailable~\cite{LaurentXMLCounting}. In this section, we show that even in that case, path summaries enable similar reasoning.

Let $op_1$, $op_2$ be two operators, such that $op_1.X$ and $op_2.Y$ contain structural identifiers. The outputs of $op_1$ and $op_2$ are ordered by the document order reflected by $X$, resp. $Y$, and are assumed duplicate-free. Assume we need to find the $op_2.Y$ IDs that are descendents of some $op_1.X$ IDs. If an ID $y_0$ from $op_2.Y$ has two different ancestors in $op_1.X$, the result of the structural join $op_1 \join op_2$ will contain $y_0$ twice, and (depending on the physical algorithm employed~\cite{StructJoins}) the output may not be in document order, requring explicit Sort and duplicate-elimination.

Path annotations provide a sufficient condition which ensures every $op_2.Y$ ID has at most one ancestor in $op_1.X$: 
{\em for any two possible paths $p_1$, $p_2$ for element IDs in $op_1.X$, $p_1$ is not an ancestor $p_2$}.
 Then, duplicate elimination and sort may be skipped, reducing the cost of the physical plan. 
For example, none of the structural joins in Figure~\ref{fig:finalQEP} require ordering or duplicate elimination.

Notice that path partitioning (Section~\ref{sec:pp}) leads to structures which naturally fulfill this requirement. If all element IDs are stored together~\cite{Grust,XRel}, or partitioned by the tags~\cite{TimberVLDBJ}, this is not the case. For instance, if all IDs of \textsf{\small parlist} elements are stored together in the document in Figure~\ref{fig:XMark}, this includes elements from paths 20 and 23, and 23 is a descendent of 20. Thus, when evaluating e.g. \textsf{\small //parlist//listitem}, some \textsf{\small listitem} IDs will get multiplied by their ancestors, requiring a duplicate elimination.

\begin{figure}[t!]
\begin{center}
\includegraphics[width=9cm]{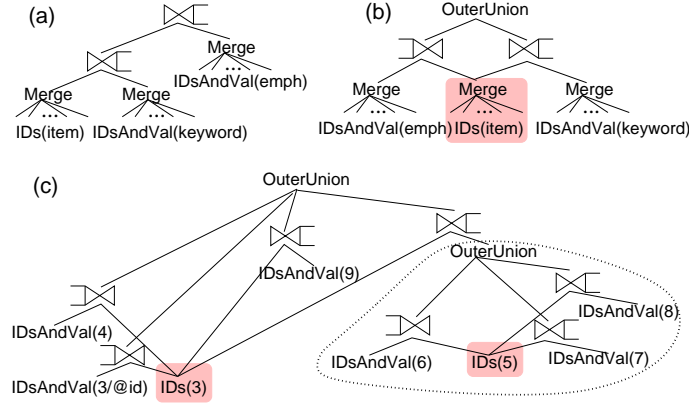}
\caption{Sample outer-union QEPs with structural joins.\label{fig:mvd}}
\end{center}
\end{figure}

\subsection{Reconstructing XML elements}
\label{sec:reconstruct}
The biggest performance issues regarding a path-partitioned store are connected to the task of reconstructing complex XML subtrees, since the data has been partitioned vertically. In this section, we study algorithms for gathering and gluing together data from multiple paths when building XML output.

A first approach is to adapt the SortedOuterUnion~\cite{JayVLDB2001} method for exporting relational data in XML, to a path-partitioned setting with structural IDs. The plan in Figure~\ref{fig:finalQEP} does just this: the components of the result (\textsf{\small name} and \textsf{\small emph} elements) are gathered via two successive structural outerjoins. In general, the plan may be more complex.
For instance, consider the query:

\begin{center}
\begin{tabular}{l}
\textsf{\small for \$x in //item return $<$res$>$ \{\$x//keyword\} \{\$x//emph\} $<$/res$>$}\\
\end{tabular}
\end{center} 

The plan in Figure~\ref{fig:mvd}(a) cannot be used for this query, because it introduces multi-valued dependencies~\cite{Ullman}: it multiplies all \textsf{\small emph} elements by all their \textsf{\small keyword} cousins, while the query asks the for \textsf{\small keyword} and \textsf{\small emph} descendents of a given item to be concatened (not joined among themselves). The plan in Figure~\ref{fig:mvd}(b) solves this problem, however, it requires materializing the \textsf{\small item} identifiers (highlighted in grey), to feed them as inputs in two separate joins. 

If the materialization is done on disk, it breaks the execution pipeline, and slows down the evaluation. If it is done in memory, the execution will likely be faster, but complex plans end up requiring more and more materialization. For instance, the simple query \textsf{\small //person} leads to the plan in Figure~\ref{fig:mvd}(c), where the IDs on both paths 3 (\textsf{\small person}) and 5 (\textsf{\small address}) need to be materialized to avoid erroneous multiplication of their descendants by successive joins. The sub-plan surrounded by a dotted line reconstructs \textsf{\small address} elements, based on \textsf{\small city}, \textsf{\small country} and \textsf{\small street}. The complete plan puts back together all components of \textsf{\small person}. 

The I/O complexity of this method is driven by the number of intermediary materialization steps and the size of the materialized results. Elements from some path $x$ must be materialized, as soon as they must be combined with multiple children, and at least one of children paths $y$ of $x$ is not annotated with \textsf{\small 1} (thus, elements on path $x$ may have zero or more children on path $y$). Some IDs are be materialized multiple times, after joins with descendent IDs at increasing nesting level. For instance, in Figure~\ref{fig:mvd}, \textsf{\small person} IDs are materialized once, and then a second time after being joined with \textsf{\small address} IDs. 
In the worst case, assuming IDs on all paths in the subtree to be reconstructed must be materialized on disk, this leads to $O(N*h/B)$ I/O complexity, where $B$ is the blocking factor. If in-memory materialization is used, the memory consumption is in $O(N*h)$. The time complexity is also $O(N*h)$.

\begin{figure}[t!]
\begin{center}
\includegraphics[width=8cm]{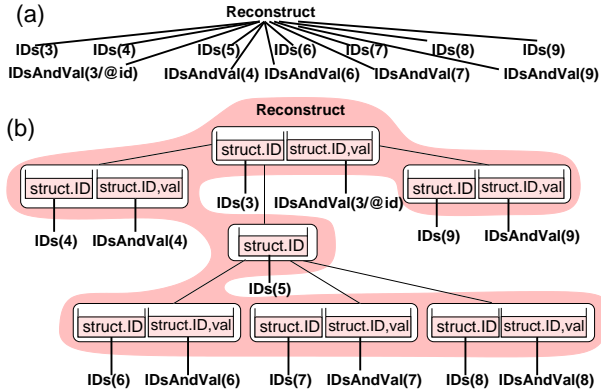}
\caption{Reconstruct plan for \textsf{\small //person} on XMark data.\label{fig:reconstruct}}
\end{center}
\end{figure}

~\\
To reduce the space requirements, we devise a second method, named Reconstruct. The idea is to read in parallel the ordered sequences of structural IDs and (ID, value) pairs from all the paths to recombine, and to produce directly textual output in which XML markup (tags) and values taken from the inputs, are concatenated in the right order.  The Reconstruct takes this order information:
\begin{itemize}
\item {\em From the path summary}: children elements must be nested inside parent elements. Thus, a \textsf{\small $<$person$>$} tag must be output (and a \textsf{\small person} ID read from IDs(3)) before the \textsf{\small $<$name$>$} child of that person, and a \textsf{\small $<$/name$>$} tag must be output (thus, all values from IDsAndVal(4) must have been read and copied) before the \textsf{\small $<$/person$>$} tag can be output.
\item {\em From the structural IDs themselves}: after an opening \textsf{\small $<$person$>$} tag, the first child of \textsf{\small person} to be reconstructed in the output comes from the path $n$, such that next structural ID in the stream IDs($n$) is the smallest among all structural ID streams corresponding to children of \textsf{\small person} elements.
\end{itemize}

Figure~\ref{fig:reconstruct}(a) outlines a Reconstruct-based plan, and Figure~\ref{fig:reconstruct}(b) zooms in into the Reconstruct itself (the shaded area). Reconstruct uses one buffer slot to store the current structural ID, and the current (ID, value) pair, from every path which contributes some data to the output. The IDs are used to dictate output order, as explained above; the values are actually output, properly nested into markup. The buffers are connected by thin lines; their interconnections repeat exactly the path summary tree rooted at \textsf{\small person} in Figure~\ref{fig:XMark}. 

A big advantage of Reconstruct is that {\em it does not build intermediary results}, thus it has a smaller memory footprint than the SortedOuterUnion approach. Contrast the QEPs in Figure~\ref{fig:mvd}(b) and Figure~\ref{fig:reconstruct}(b): the former needs to build \textsf{\small address} elements separatedly, while the latter combines all pieces of content directly. A second advantage is that the Reconstruct is pipelined, unlike the SortedOuterUnion, which materializes \textsf{\small person} and \textsf{\small address} IDs.

The Reconstruct has $O(N)$ time complexity. It needs one buffer page to read from every path which contributs some data to the output, thus it has $O(n)$ memory needs, where $n$ is the number of paths from which data is combined; especially for large documents, $n \ll N*h/B$, thus the Reconstruct is much more memory-efficient than the SortedOuterUnion approach.

\section{Experimental evaluation}
\label{sec:experiments}
We have implemented path summaries within the XQueC path-partitioned system~\cite{XQueCVLDB2003,XQueCEDBT2004}, 
and as an independent library~\cite{XSum}. This section describes our experience with building and exploiting summaries, alone or in conjunction with a path-partitioned store. 

We use the documents from Table~\ref{tab:PS}, ranging from 7.5 MB to 280 MB, with relatively simple (Shakespeare) to extremely complex (TreeBank) structure. Experiments are carried on a Latitude D800 laptop, with a 1.4 GHz processor, 1 GB RAM, running RedHat 9.0. We use XQueC's path-partitioned storage system~\cite{XQueCVLDB2003}, developed based on the popular persistent storage library BerkeleyDB from \textsf{\small www.sleepycat.com}. The store uses B+-trees, and provides efficient access to the IDs, or (ID,val) pairs, from a given path, in document order. All our development is Java-based; we use the Java HotSpot VM 1.5.0. All times are averaged over 5 runs.

\subsection{Path summary size and serialization}
Summary sizes, in terms of nodes, have been listed in Table~\ref{tab:PS}. We now consider the sizes attained by serialized stored summaries. Two choices must be made: ($i$)~XML or binary serialization, ($ii$)~direct or precomputed encoding of parent-child cardinalities (Section~\ref{sec:summaries}), for a total of four options.
XML serialization is useful since summaries may be easily inspected by the user, e.g. in a browser. Summary nodes are serialized as elements, and their annotations as attributes with 1-character names. Binary serialization yields more compact summaries; summary node names are dictionary-encoded, summary nodes and their labels are encoded at byte level. Pre-computed serialization is more verbose than the direct one, since \textsf{\small n1} and \textsf{\small n+} labels may occupy more than \textsf{\small 1} and \textsf{\small +} labels. 

Table~\ref{tab:sum2} shows the {\em smallest} serialized summary sizes (binary with direct encoding). Properly encoded, information-rich summaries are much smaller than the document: 2 to 6 orders of magnitude smaller, even for the large TreeBank summary (recall Table~\ref{tab:PS}).

\begin{table}[t!]
\begin{center}
{\small \begin{tabular}{|l|r|r|r|}
\hline
Doc. & Shakespeare & XMark$11$ & XMark$233$  \\
\hline
Size (MB)& 7.5 & 11 & 233  \\
XML pre-comp (KB)& 0.68 & 4.85 & 4.95 \\
XML pre-comp / size&8*10$^{-5}$&4*10$^{-4}$&2*10$^{-5}$ \\
\hline
Doc. & SwissProt & DBLP 2005& TreeBank \\
\hline
Size (MB) & 109 & 280 & 82 \\
XML pre-comp (KB) & 3.11 & 1.62 & 2318.01 \\
XML pre-comp / size & 2*10$^{-5}$&5*10$^{-6}$& 3*10$^{-2}$\\
\hline
\end{tabular}}
\end{center}
\caption{Serialized summary sizes (binary, direct encoding).\label{tab:sum2}}
\end{table}

We also measured XML-based summary encodings for the documents in Table~\ref{tab:sum2}, and found they are 2 to 5 times larger than the direct binary one. We also measured the size of the binary pre-computed summaries, and found it always within a factor of 1.5 of the direct binary one, which is quite compact.

\subsection{Relevant path computation}
We now study the performance of the relevant path set computation algorithm from Section~\ref{sec:computingPaths}. We use the XMark233 summary as representative of the moderate-sized ones, and Treebank as the largest (see Table~\ref{tab:sum2}).
Table~\ref{tab:xmarkPaths} shows the time needed to compute the relevant path sets for the 20 queries of the XMark benchmark~\cite{XMark}, on the XMark111 summary, serialized in binary format with pre-computed information. The query patterns have between 5 and 18 nodes. Path computation is very fast, and takes less than 50 ms, demonstrating its scalability with complex queries. 

\begin{table}
\begin{center}
{\small \begin{tabular}{||r|r|r|r|r|r|r|r|r|r|r||}
\hline
\hline
query no.&1&2&3&4&5&6&7&8&9&10\\
\hline
time (ms)&14&14&14&15&14&14&14&29&46&29\\
\hline
\hline
query no.&11&12&13&14&15&16&17&18&19&20\\
\hline
time (ms)&28&28&14&14&15&16&15&14&15&14\\
\hline
\hline
\end{tabular}}
\end{center}
\caption{Computing relevant paths for the XMark queries.\label{tab:xmarkPaths}}
\end{table}

We now measure the impact of the serialization format on the relevant path computation time. The following table shows this time for the XMark queries 1 and 9, for which Table~\ref{tab:xmarkPaths} has shown path computation is fastest, resp. slowest (times in milliseconds):

\begin{center}
\begin{tabular}{|c|r|r|r|r|}
\hline
query no. & XML dir. & XML pre-cp. & bin. dir. & bin. pre-cp \\
\hline
1 & 73.0 & 37.0 & 22.3 &14.2\\
\hline
9 & 255.7 & 133.6 & 98.6 & 46.4\\
\hline
\end{tabular}
\end{center}

Path computation on an XML-ized summary is about 4-5 times slower than on the binary format, reflecting the impact of the time to read the summary itself. Also, the running time on a pre-computed summary is about half of the running time on a direct-encoded one. This is because with direct encoding, path set minimization requires a second summary traversal, as explained in Section~\ref{sec:associating}. The space saving of the binary, direct encoding over the binary pre-computed encoding (less than 50\%) is overcome by the penalty direct encoding brings during relevant path sets computations. We thus conclude the {\em binary, pre-computed encoding} offers the best time-space compromise, and will focus on this only from now on. If the optimizer caches query plans, however, the binary direct encoding may be preferrable.

\begin{table}[t!]
\begin{center}
\begin{tabular}{l}
{\small \textbf{TK$n$}:~\textsf{//S/VP/(NP/PP)$^n$/NP}} $\quad$ {\small \textbf{T0}:~\textsf{//A}} $\quad$ {\small \textbf{T1}:~\textsf{//NP}} $\quad$ {\small \textbf{T2}:~\textsf{//NNP}}\\
{\small \textbf{T3}:~\textsf{//WHADVP}} $\quad$ {\small \textbf{T4}:~\textsf{//NP//NNP}} $\quad$
{\small \textbf{T5}:~\textsf{//S[NPP][\_COMMA\_]/PP}}\\
{\small \textbf{T6}:~\textsf{//ADJP/PP/NP}} $\qquad\quad$
{\small \textbf{T7}:~\textsf{/FILE/EMPTY/S[VP/S]/NP/VP}}\\
\end{tabular}
\caption{XPath renditions of query patterns on TreeBank data.\label{tab:tbQueries}}
\end{center}
\end{table}

\begin{figure}[t!]
\begin{center}
\includegraphics[width=5cm,height=3cm]{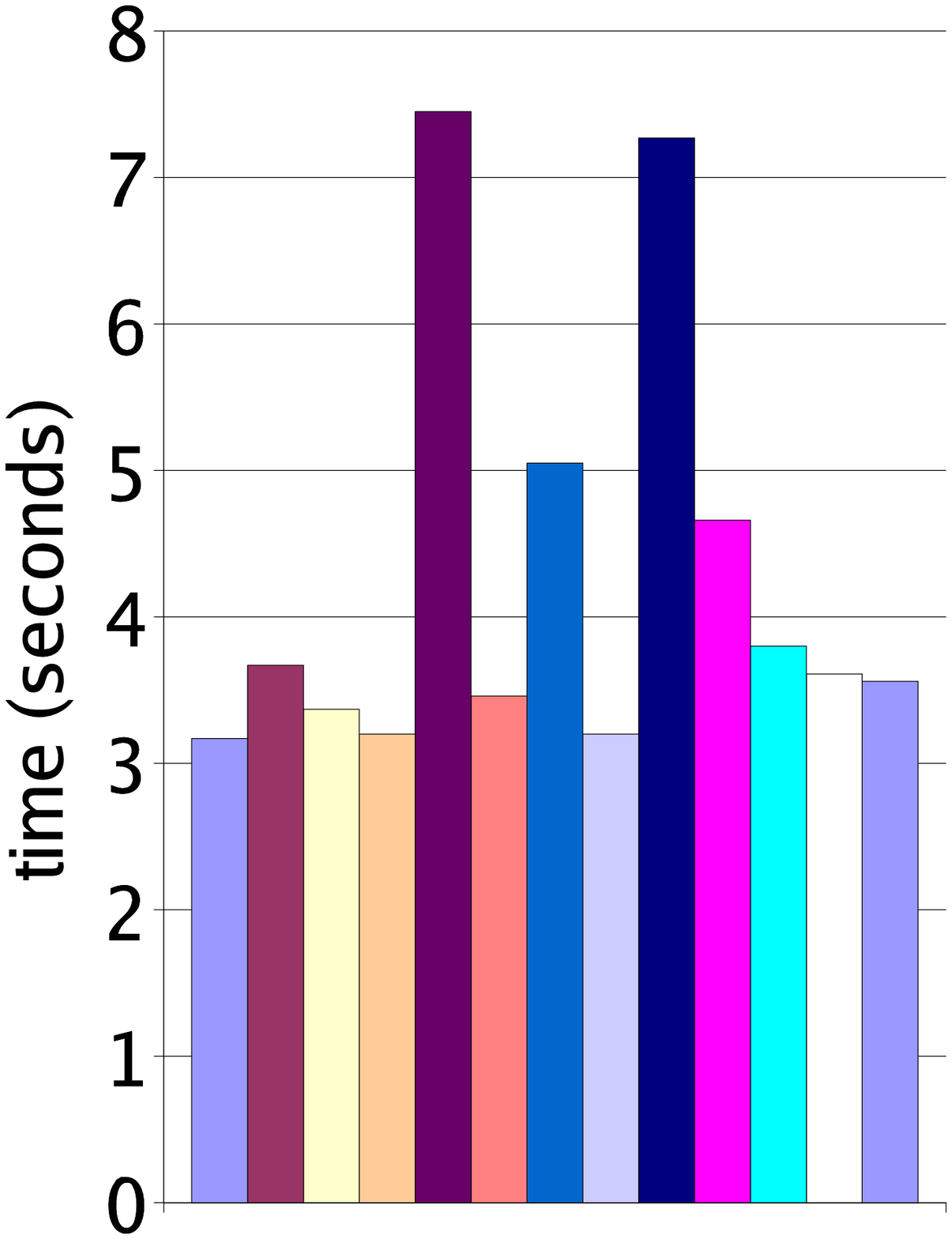}\hspace{-2.2cm}
\includegraphics[width=4.5cm,height=3cm]{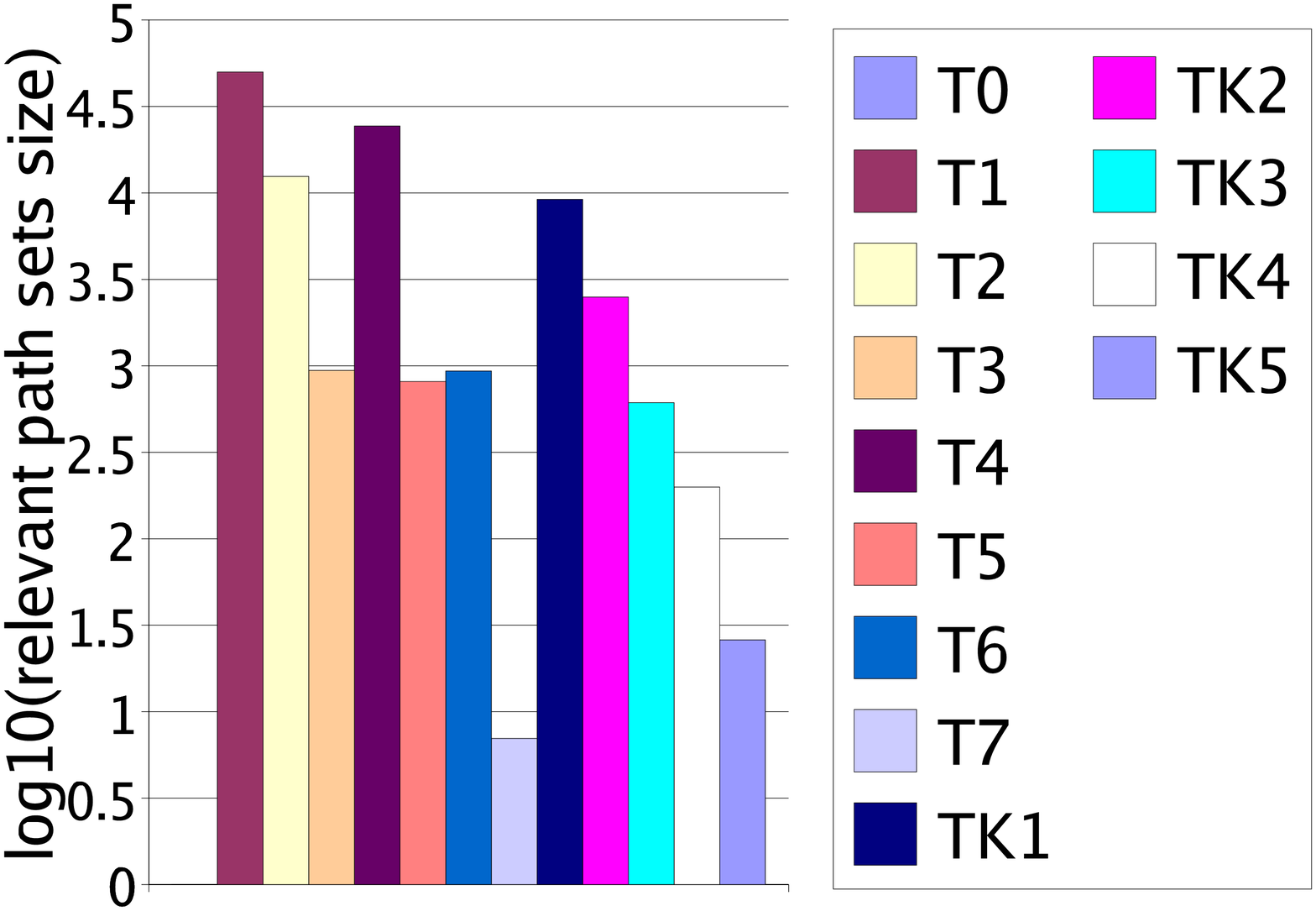}
\caption{Relevant path computation times on TreeBank (left) and resulting relevant path set size (right, log scale).\label{fig:pathsTreebank}}
\end{center}
\end{figure}

We now consider the TreeBank summary (in binary pre-computed encoding), and a set of query patterns, shown in Table~\ref{tab:tbQueries} as XPath queries for simplicity (however, unlike XPath, we compute relevant path sets for {\em all} query nodes). Treebank tags denote parts of speech, such as \textsf{\small S} for \underline{s}entence, \textsf{\small VP} for \underline{v}erb \underline{p}hrase, \textsf{\small NP} for \underline{n}oun \underline{p}hrase etc. TK$n$ denotes a parameterized family of queries taken from~\cite{koch}, where the steps \textsf{\small /NP/PP} are repeated $n$ times. 

Figure~\ref{fig:pathsTreebank} (left) shows the times to compute the relevant paths for these queries. Due to the very large summary (2.3 MB), the times are measured in seconds, two orders of magnitude above those we registered for XMark.
Queries T0 to T3 search for a single tag. The time for T0 is spent traversing the summary only, 
since the tag \textsf{\small A} is not present in the summary,\footnote{A tag dictionary at the beginning of the summary allows detecting erroneous tags directly. We disabled this feature for this measure.} thus no stack entries are built. The other times can be decomposed into: the constant summary traversal time, equal to the time for T0; and the time needed to build, check, and prune stack entries. 

T1 takes slightly more than T2, which takes more than T3, which is very close to T0. The reason can be seen by considering at right in Figure~\ref{fig:pathsTreebank} the respective number of paths: T1 results in much more paths (about 50.000) than T2 (about 10.000) or T3 (about 1.000). More relevant paths means more entries to handle. 

The time for T4 is the highest, since there are many relevant paths for both nodes. Furthermore, an entry is created for all \textsf{\small NP} summary nodes, but many such entries are discarded due to the lack of \textsf{\small NNP} descendents.  T5, T6 and T7 are some larger queries; T6 creates some \textsf{\small ADJ} entries which are discarded later, thus its relatively higher time. The times for TK$n$ queries {\em decreases as $n$ \em increases}, a tendency correlated with the number of resulting paths, at right in Figure~\ref{fig:pathsTreebank}. Large $n$ values mean more and more selective queries, thus entries in the stacks of nodes towards the beginning of the query (\textsf{\small S}, \textsf{\small VP}) will be pruned due to their lack of required descendents (\textsf{\small NP} and \textsf{\small PP} in the last positions in the query).

The {\em selfparent} encoding proved very useful for queries like T4. For this query, we counted more than 75.000 relevant path pairs (one path for \textsf{\small NP}, one for \textsf{\small NPP}), while with the {\em selfparent} encoding, only 24.000 stack entries are used. This demonstrates the interest of {\em selfparent} pointers in cases where there are many relevant paths, due to a large summary and/or to \textsf{\small *} query nodes.

\begin{figure}
\begin{center}
\begin{minipage}{8.4cm}
\includegraphics[width=4.4cm,height=3.1cm]{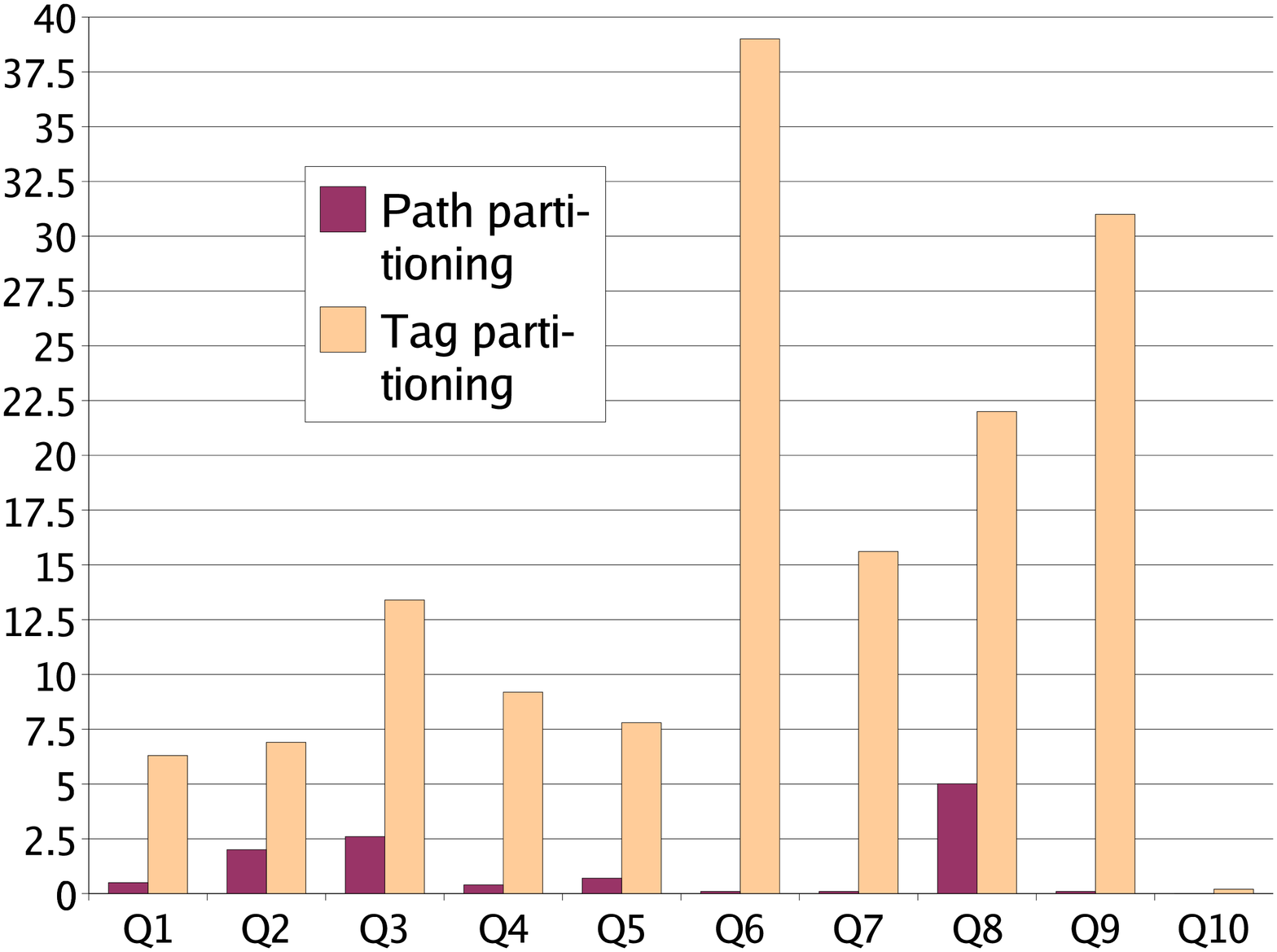}
\end{minipage}
\begin{minipage}{3cm}
{\small \begin{tabular}{ll}
Q1&\textsf{//europe//item/descr}\\
Q2&\textsf{//regions//item//descr}\\
Q3&\textsf{//europe//parlist//bold}\\
Q4&\textsf{//europe//parlist//listitem}\\
Q5&\textsf{//item//descr//keyword}\\
Q6&\textsf{//Entry//METAL//Descr}\\
Q7&\textsf{//categ//listitem//text}\\
Q8&\textsf{//parlist//listitem//text}\\
Q9&\textsf{//dblp//book//author}\\
Q10&\textsf{//dblp//book//title}
\end{tabular}}
\end{minipage}\\
\begin{minipage}{6cm}
\includegraphics[width=6cm,height=3.1cm]{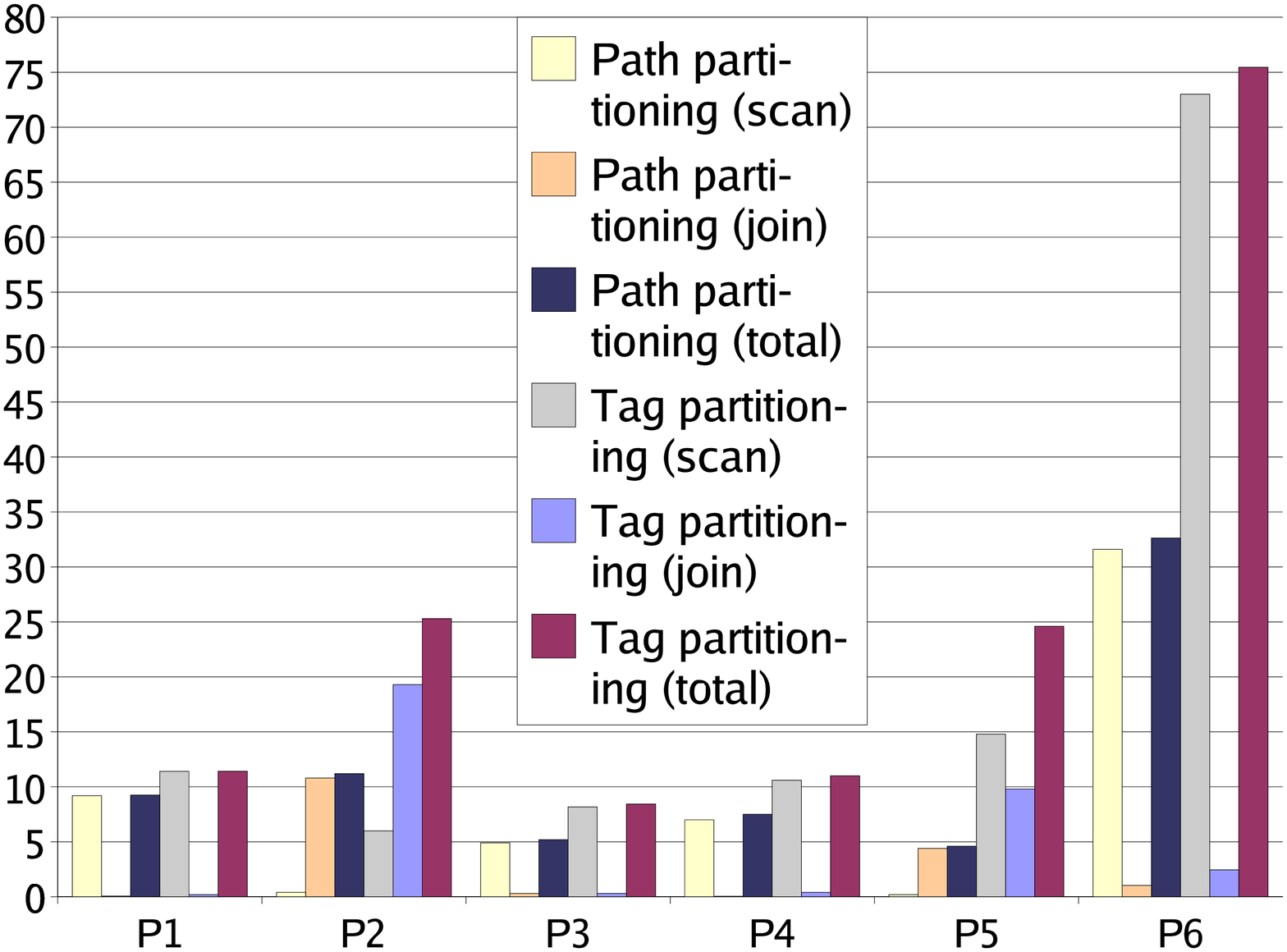}
\end{minipage}
\hspace{3mm}
\begin{minipage}{7.8cm}
{\small \begin{tabular}{ll}
P1&\textsf{for \$x in //person, \$y in \$x//name, \$z in \$x//watch}\\
P2&\textsf{for \$x in //asia, \$y in \$x//item, \$z in \$y//description}\\
P3&\textsf{for \$x in //item, \$y in \$x//description, \$z in \$x//parlist}\\
P4&\textsf{for \$x in //item, \$y in \$x//parlist, \$z in \$x//keyword}\\
P5&\textsf{for \$x in //categories, \$y in \$x//parlist, \$z in \$x//text}\\
P6&\textsf{for \$x in //article, \$y in \$x//title, \$z in \$x//year}\\
\end{tabular}}
\end{minipage}
\caption{Binding variables with path- and tag-partitioning.\label{fig:bind}}
\hfill
\end{center}
\end{figure}

\subsection{Variable binding with path partitioning}
\label{sec:exp-binding}
We measured the time needed to {\em bind} variables to element IDs on our path partitioned store. We identify relevant paths based on the summary, read the ID sequences for relevant paths, and perform structural joins if needed. For comparison, we also implemented a similar store, but where IDs are partitioned {\em by their tags}, not by their paths, as in~\cite{MixedMode,TimberVLDBJ}. On both stores, the StackTreeDesc~\cite{StructJoins} structural algorithm was used to combine structural IDs. 

Figure~\ref{fig:bind} shows the execution times for 10 XPath queries (Q6 on SwissProt, Q9 and Q10 on DBLP, the others on a 116 MB XMark document), and 6 tree patterns (P1 to P6 on the 116 MB XMark). In Figure~\ref{fig:bind}, path partitioning takes advantage of the relevant path computation to achieve tremendous performance advantages (up to a factor of 400~!) over tag partitioning. This is because often, many paths in a document end in the same tag, yet only a few of these paths are relevant to a query, and our relevant path computation algorithm identifies them precisely. For the patterns P1 to P6, we split the binding time in ID scan, and ID structural join. We see that the performance gain of path partitioning comes from its reduced scan time. For all these queries, the relevant path computation time was less than $20$ ms, thus 2 to 6 orders of magnitude less than execution time. This confirms the interest of the access method selection algorithm enabled by the summary. 


\paragraph*{Measuring the difference between path and tag partitioning} A legitimate question arises: do tag and path partitioning differ significantly in general, or is the difference only noticeable in some contrived data sets and queries~?

The answer depends on the number of different paths in document $d$ which end in a given tag $t$; we call this number the {\em fan-in of $t$ in $d$} and denote it $fin_{t,d}$. If the fan-in of all tags in a given document is $1$, then tag partitioning and path partitioning coincide. The bigger the fan-in is, the more likely it is that tag partitioning and path partitioning will perform differently when binding variables.

Considering the maximum (or the average) value of $fin_{t,d}$, over all tags $t$ in a document $d$, does not account for the relative importance of each tag $t$ in $d$. Thus, we consider the {\em median fan-in of $d$}, defined as:
\[mf_d = \sum_{t\in d}fin_{t,d}*N_{t,d}/N_d\]
where $N_d$ is the number of (element and attribute) nodes in $d$, while $N_{t,d}$ is the number of nodes labeled $t$. This measure gives greater weight to tags well-represented in the document, since they can be seen as ``more representative''. The results for the documents in Table~\ref{tab:PS} are shown in Table~\ref{tab:fanin}.

\begin{table}[t]
{\footnotesize \begin{center}
\begin{tabular}{||c|r|r|r|r|r|r|r|r||}
\hline
\hline
Doc. & TreeBank & INEX & XMark$111$ & SwissProt & Shakespeare & DBLP & NASA & UW \\
\hline
recursion & yes & yes & yes & no & no & no & yes & no\\
\hline
$max f_{t,d}$ &$49,901$ &$1,722$ &$99$ & $39$ &$9$ &$8$ &$9$ & $1$ \\
\hline
$mf_d$ &$19,187.82$  & $485.43$ &$15.79$ & $11.08$&$6.06$ &$5.47$ &  $2.92$& $1.0$\\
\hline
\hline
\end{tabular}
\end{center}}
\caption{Median fan-in of different XML documents.\label{tab:fanin}}
\end{table}

In  all but the UW course data, $mf_{d}$ is quite important. In the UW course datasets, the maximum fan-in registered is $1$, thus tag partitioning and path partitioning coincide.  Interestingly, this document has been produced from a {\em relational} data set, by a database Ph.D. student~\cite{UWXML}, who might have thought it proper to give distinct names only; this cannot be expected in general.
The $maxf_{t,d}$ and $mf_{d}$ values may be quite different,
which justifies introducing the $mf_{d}$ measure.
Let us analyze the possible reasons for large fan-ins.

{\em Recursive elements}, like XMark's {\small \textsf{parlist}}, are one source, since the recursive element tag may appear on various paths. Recursion is actually encountered in a significant part of XML documents on the Web~\cite{LaurentXMLCounting}.

Another source is the presence of {\em common tags} like {\small\textsf{name}}, {\small\textsf{text}}, {\small\textsf{descr}}, {\small\textsf{@from}}, which apply in different contexts, and thus appear on different paths. Part of the meaning of such data resides in its path. For example, in SwissProt, {\small\textsf{descr}} elements appear under paths like {\small\textsf{root/entries/PROTEIN}} and {\small\textsf{/root/entries/METAL}}, making them relevant for different queries. 

A third source is XML {\em flexibility}, which may lead to a data annotation at the level of tags (metadata), or at the level of the values (data). For example, the XMark {\small\textsf{item}} data is split on 6 different paths, under the elements {\small\textsf{europe}}, {\small\textsf{asia}} etc. 

Finally, {\em textual} XML documents tend to exhibit important fan-in values. Tags in such documents correspond to language components (e.g., ``verb group'', ``nominal group'', ``attribute'', ``phrase'') or markup elements (e.g., ``italic'', ``bold''), which can be arbitrarily nested. This phenomenon is present to some extent in the XMark documents, and is very visible in TreeBank and INEX. Such documents are representative for an important class of text-centered XML applications, arguably, of more interest for XML research than XML-ized relational sources. Variable binding based on path partitioning is likely to outperform tag partitioning on such data.


\paragraph*{Impact of path minimization} Relevant path computation finds that the second tag in Q1-Q5 is useless (Section~\ref{sec:associating}), thus IDs for those tags are not read, in the measures in Figure~\ref{fig:bind}. Turning minimization off increased the running time by 15\% to 45\%.

\begin{figure}
\begin{center}
\includegraphics[width=5.8cm,height=3.4cm]{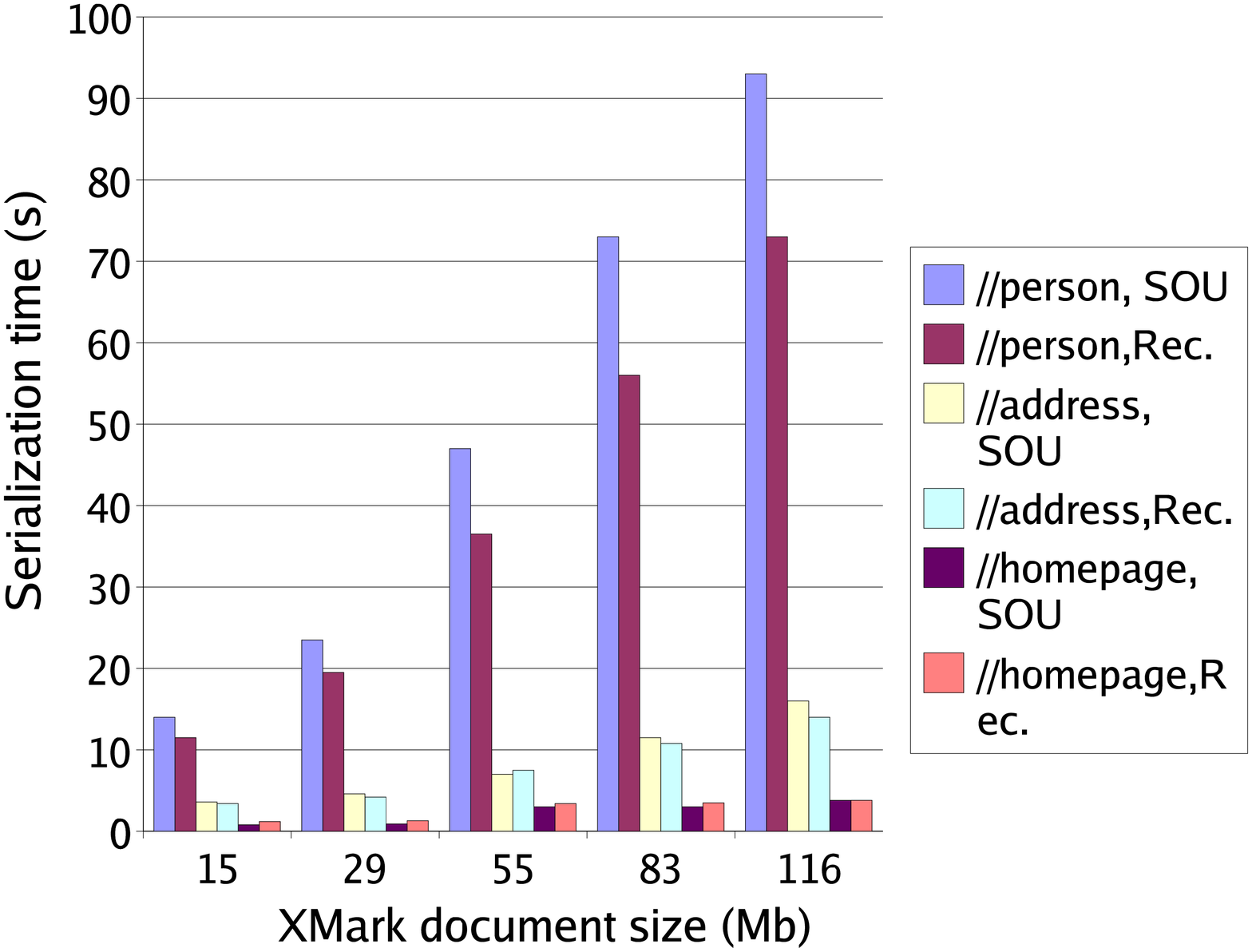}
\caption{SortedOuterUnion and Reconstruct performance.\label{fig:exp-reconstruct}}
\end{center}
\end{figure}

\subsection{Reconstructing path-partitioned data}
\label{sec:reconstruction}
We tested the performance of the two document reconstruction methods described in Section~\ref{sec:reconstruct}, on our path-partitioned store. Figure~\ref{fig:exp-reconstruct} shows the time to build the full serialized result of \textsf{\small //person}, \textsf{\small //address}, \textsf{\small //homepage}, on XMark documents of increasing sizes. The sorted outer union (denoted SOU in Figure~\ref{fig:exp-reconstruct}) materialized intermediary results in memory. On the XMark116 document, \textsf{\small //person} outputs about 15 MB of result. As predicted in Section~\ref{sec:reconstruct}, both methods scale up linearly. The Reconstruct is noticeably faster when building complex elements such as \textsf{\small address} and \textsf{\small person}. Furthermore, as explained in Section~\ref{sec:reconstruct}, it uses much less memory, making it interesting for a multi-user, multi-query setting. 

\subsection{Conclusions of the experiments}
Our experiments have shown that path summaries can be serialized very compactly;  the binary encoded approach yields the best trade-off between compactness and relevant path computation performance. Our path computation algorithm has robust performance, and produces intelligently-encoded results, even for very complex summaries. Path partitioning takes maximum advantage of summaries; used in conjunction with structural identifiers and efficient structural joins, it provides for very selective access methods. Scalable reconstruction methods make path partitioning an interesting idea in the context of current-day XML databases.

%
\section{Related work}
Path summaries and path partitioning are not new~\cite{AAN01,ToX,goldman97,XParent,paperDataX2004,PIndex,XRel}. Other more elaborate structure indexes have been proposed~\cite{FBIndex,PIndex}, however, they are very complex to build and to maintain, and thus are built for a few selected paths only.
Complex, richer XML summaries have also been used for data statistics;they tend to grow large, thus only very small subsets are kept~\cite{Garofalakis}. This is appropriate for cardinality estimation, yet inadequate for access method selection, since some structure information is lost.

~\\
An interesting class of compressed structure is described in~\cite{koch}, and it is used as a basis for query processing. This approach compresses the XML structure tree into a compact DAG, associating to each DAG node the set of corresponding XML element nodes. An interesting comparison is the number of nodes created in the path summary as explained in Section~\ref{sec:summaries}, denoted $|PS|$, and the number of nodes in the DAG of~\cite{koch}, denoted $\vert DAG\vert$. The results are shown in Table~\ref{tab:koch}.

\begin{table}[h]
\begin{center}
\begin{tabular}{||c|r|r|r|r|r|r|r||}
\hline
\hline
Document & Shakespeare & XMark$15$ & Nasa & Treebank & SwissProt & XMark$111$ & DBLP \\
\hline
\hline
Size & 7.5 MB & 15 MB & 24 MB & 82 MB & 109 MB & 111 MB & 128 MB \\
\hline
$|PS|$ & 58 & 511 & 24 & 338,738 & 117 & 514 & 125\\
\hline
$|DAG|$ & 1,121 & 10,629 & 8,391 & 319,654 & 38,936 & 38,655 &  326\\
\hline
\hline
\end{tabular}
\caption{Number of path summary nodes vs. the number of nodes in the DAG obtained by bisimulation.\label{tab:koch}}
\end{center}
\end{table}

Table~\ref{tab:koch} shows that a path summary is generally smaller (in some cases by two orders of magnitude) that the DAG obtained by~\cite{koch}. This is explained by the fact that for two XML nodes to correspond to a single summary node, a path summary only requires that their incoming paths be the same, whereas the DAG summary introduced in~\cite{koch} also requires their underlying structure to be similar. The difference is striking, e.g., in the case of XMark data sets; the presence of recursive, variable and repeated structure in the data leads to a relatively large DAG, but a compact path summary. Going from a $15$ MB to $111$ MB one, the path summary adds only three paths, but the DAG size has more than tripled~! For Treebank, the DAG is slightly smaller than the path summary (by less than $3\%$). We thus argue that path summaries are generally much more robust and therefore of practical interest.

In~\cite{bunemanICDE2005}, the authors propose a specialized query processing framework based on the summaries described in~\cite{koch}. The authors present an approach for handling DAG-compressed structures throughout the query processing steps, which reduces the risk that the unfolded compressed structure would outgrow the available memory. In contrast, we make the point that path summaries can be added with minimal effort into existing XQuery processing systems, and that they marry well with efficient techniques such as structural identifiers and structural joins.

~\\
Path information has been used recently for XPath materialized view-based rewriting~\cite{Balmin} and for access method selection~\cite{ULoad,Barta}.  Our work is complementary in what concerns the path summary, since we formalized and presented efficient algorithms for exploiting summaries, which could be integrated with these works. As we have demonstrated, some documents yield large summaries, whose exploitation may raise performance problems, therefore, we have provided an efficient relevant path computation algorithm, which furthermore performs some interesting query minimizations. The only previous path summary exploitation algorithm concerns simple linear XPath path queries only~\cite{AAN01}, and it does not perform any minimization. With respect to~\cite{ULoad,Barta}, in this work we focused on formalizing relevant path computation, and showing its benefits in the particular context of a path-partitioned store.

~\\
Many works target specifically query minimization, sometimes based on constraints, e.g.~\cite{SihemMinimizing,NEXT,LaksRecent}. We show how summaries can be used as practical structures encapsulating constraints.
\begin{itemize}
\item Some of the benefits offered by a summary can also be attained by using DTD or XML Schema information. However, a large part of the XML existing body of documents may lack a schema (in the study~\cite{LaurentXMLCounting}, 40\% of the documents had a DTD, and less than 10\% had an XML Schema). Even in the absence of schema information, a summary is very easy to build and to exploit, as soon as one has had a chance to look at the data (which is the case in any persistent database, since the data has been loaded).
\item Some of the summary benefits cannot be attained by using schemas, because the summary is more precise in some aspects, such as the actual depth of recursive elements etc. 
\end{itemize}
Constraint-independent minimization techniques are orthogonal to our work and can be successfully combined. 

~\\
With respect to path partitioning, we considered  the task of retrieving IDs satisfying given path constraints as in~\cite{Balmin,Barta,PIndex} and shown that structural IDs and joins efficiently combine with path information. Differently from~\cite{Balmin,Barta,PIndex} which assume available a persistent tree structure, we also considered the difficult task of re-building XML subtrees from a path-partitioned store. We studied an extension of an existing method, and proposed a new one, faster and with much lower memory needs.

The starting point of this work is the XQueC compressed XML prototype~\cite{XQueCVLDB2003,XQueCEDBT2004}. The contributions of this paper on building and exploiting summaries for optimization have a different scope. An early version of this work has been presented in an informal setting, within the French database community only~\cite{PathSequences}. A 2-pages poster based on this work is currently under submission.

\section{Conclusion and perspectives}
We have described a practical approach for building and exploiting path summaries as metadata in a persistent XML repository, i.e., information about the structure encountered in the XML document. We have shown how summaries can be combined with path partitioning to achieve efficient, selective data access, a plus for processing queries with a complex navigation requirements.

Our own experience developing the summary was first included in our XQueC~\cite{XQueCVLDB2003,XQueCEDBT2004} XML compression project. Subsequently, we isolated it out of the XQueC prototype, and found it useful applications, which we briefly describe below; the prototype is freely available~\cite{XSum}. 

\paragraph*{Apprehending varied-structure data sources} In the framework of the INEX\footnote{INEX stands for Initiative for the Evaluation of XML Information Retrieval; see http://inex.is.informatik.uni-duisburg.de.} collaborative effort, we concentrated on designing an integrated conceptual model out of heterogeneously-structured bibliographic data sources. As a side effect of building summaries, XSum also generates image files of such summaries~\cite{XSum}. We used this feature to get acquainted to the sources and visualize their structure. This is in keeping with the initial Dataguide philosophy of using summaries for exploring data sets~\cite{goldman97}.

\paragraph*{Physical data independence} We developed a materialized view management tool for XQuery, called ULoad~\cite{ULoad}. This tool includes a query rewriting module based on views, which naturally leads to containment and equivalence problems. ULoad judges containment and equivalence {\em under summary constraints}, thus exploiting summaries and path annotations.

\paragraph*{Query unfolding} An ongoing work in the Gemo group requires a specific form of query unfolding. As soon as an XQuery returns some elements found by some unspecified navigation path in the input document (that is, using the descendant axis), the query must be rewritten so that it returns {\em all elements on the path from the document root to the returned node}, not just the returned node as regular XPath semantics requires. For instance, the query \textsf{\small //person} in an XMark document must be transformed into the query in Table~\ref{tab:unfoldedQuery}. A summary is an useful tool in this context.

\begin{table}[ht]
{\small \begin{tabular}{|l|l|}
\hline
&\textsf{\small for \$x1 in document(``xmark.xml'')/site}\\
&\textsf{\small return $<$site$>$ \{ for \$x2 in \$x1/people} \\
\textsf{document(``xmark.xml'')//person}&$\qquad\qquad\qquad\quad$\textsf{\small return $<$people$>$ for \$x3 in \$x2/person return \$x3}\\
&$\qquad\qquad\qquad\qquad\qquad$\textsf{\small $<$/people$>$ \}}\\
&$\qquad\quad$\textsf{\small $<$/site$>$}\\
\hline
\end{tabular}}
\caption{Sample summary-based query unfolding.\label{tab:unfoldedQuery}}
\end{table}

\paragraph*{Perspectives} Our ongoing work focuses on adding to the XSum library a version of the containment and equivalence algorithms implemented in ULoad. We are also considering the joint usage of summary and schema information for XML tree pattern query rewriting and containment; we anticipate that this combined usage provides increased information and thus more opportunities for optimization.

We are also currently extending ULoad to support XQuery updates; accordingly, we expect to implement summary maintenance under data modifications in XSum. It is to be noted that summary maintenance has very low complexity, using our notion of summary~\cite{goldman97}, thus we do not expect this to raise difficult issues. 

\paragraph*{Acknowledgements} The authors are grateful to Christoph Koch for providing us with his XML compressor code~\cite{koch}, and to Pierre Senellart for sharing with us his query unfolding application.

\footnotesize{
\bibliographystyle{plain}
\bibliography{abmp}}

\end{document}